\documentclass[11pt]{article}
\usepackage{amsmath,euscript,amssymb}
\usepackage[T2A]{fontenc}
\usepackage[cp866]{inputenc}
\usepackage[dvips]{graphicx}
\def\v1{\vspace{1cm}}
\def\be{\begin{equation}}
\def\ee{\end{equation}}
\def\bc{\begin{center}}
\def\ec{\end{center}}

\newcommand{\bea}{\begin{eqnarray}}
\newcommand{\eea}{\end{eqnarray}}

\begin{document}
\title{
Solution of Constraints in Theory of  Relativistic String
 }
 \author{ B.M. Barbashov${}^{1}$ and  V.N. Pervushin${}^{1}$
 \\[0.3cm]
{\normalsize\it $^1$ Joint Institute for Nuclear Research},\\
 {\normalsize\it 141980, Dubna, Russia} 
}

\date{\empty}
\maketitle
\medskip

\begin{abstract}
 {\noindent
  The Hamiltonian theory of  a relativistic string
  is  considered in  a specific
  reference frame in terms the diffeo-invariant variables.

   The evolution parameter and energy   invariant
  with respect to the time-coordinate transformations are constructed,
   so that the
 dimension of the kinemetric  group of diffeomorphisms coincides with
 the  dimension of a set of variables whose velocities are removed
 by the Gauss-type constraints in accordance with the second N\"other theorem.
   This  coincidence
  allows us to solve the energy constraint, and
 fulfil Dirac's Hamiltonian reduction.
}
\end{abstract}

\vspace{2cm}

 \centerline{\bf Invited Talk at INTAS Summer School and Conference }

 \centerline{\bf ``New
trends in High-Energy Physics''}

 \centerline{\bf Yalta, Crimea,
Ukraine, September 10 - 17, 2005}

\newpage


\section{Introduction}

 The main difficulty of the theory of a relativistic string is
 the group of diffeomorphisms, i.e. general coordinate
 transformations \cite{1}.

  There is
  an essential difference between the frame group
 of the Lorentz -- Poincar\'e-type
  leading to a set of
 initial data
 and the {\it diffeomorphism  group} of general coordinate transformations
  restricting these initial data by {\it constraints}. This difference
  was revealed by two  N\"other theorems  \cite{Noter}\footnote{One of
  these theorems (the second) was  formulated by Hilbert in his famous paper \cite{H}
   (see also its revised version  \cite{H24}). We should like to thank
   V.V. Nesterenko who draw our attention to this fact \cite{d1986}.}.

 Starting position of the paper is the group of
 diffeomorphisms of this ``Hamiltonian frame'' known as the kinemetric
 one in general relativity \cite{vlad}. This kinemetric group contains
  global reparametrizations of the coordinate time and  local
  transformations of a spatial coordinate, and it
   requires
  to distinguish
 a set of variables with the same dimension,
 velocities (or momenta) of which
 are removed by the Gauss-type constraints
 from the phase space of  physical
 variables in  accordance
  with the second N\"other theorem.

A similar Hilbert-type \cite{H,H24} {\it geometro}-dynamic
  formulation of  special relativity (SR) \cite{bpp}
  with reparameterizations
  of the coordinate evolution parameter shows that
 the energy constraint fixes
 a velocity of one of the dynamic variables
 that becomes a  evolution parameter.
In particular, in SR such a
 dynamic evolution parameter is well known, it is the
 fourth component of the Minkowskian space-time
 vector.

 In order to realize a similar construction of
  the
  Dirac -- ADM Hamiltonian GR \cite{dir},
 one should  point out  in GR a homogeneous variable that can be
  a diffeo-invariant evolution parameter in the field space of
  events \cite{WDW}
   in accordance with the
  kinemetric diffeomorphism group of GR in the ``Hamiltonian frame''
  \cite{vlad}.

 Thus, in the standard  approach to a relativistic string
 this homogeneous evolution parameter is not split,
  so that  the
  dimension of the Hamiltonian constraints does not coincide with
  the dimension of the diffeomorphism group.

 In the present paper,
 a evolution parameter
   is identified with the homogeneous component of the time-like
   variable \cite{fr} in accordance with the second N\"other theorem.

\section{Diffeo-Invariant Content of a Relativistic String}

To illustrate the invariant Hamiltonian reduction let us consider
the action for a relativistic string \cite{1} in the form which
has been done by L. Brink, P. Di Vecchia, P. Howe \cite{2}

\be\label{ss} S=-\frac{\gamma}{2}\int\int d^2
u\sqrt{-g}g^{\alpha\beta}\partial_{\alpha}x^{\mu}
(u_0,u_1)\partial_{\beta}x_{\mu}(u_0,u_1),~~~~~~ (u_0,u_1)=(\tau,
\sigma),\ee where $x^{\mu}(\tau,\sigma)$ -- string coordinates
given in d-dimension space-time $(\mu=0,1,2,\ldots ,d-1)$,
$g_{\alpha\beta}(u_0,u_1)$ -- is a second-rank metric tensor on
the string surface (two dimensional Riemannian space $u_0,u_1$).

Now let us consider the Hamiltonian scheme which is based on the
Arnowitt--Deser--Misner parametrization of metric tensor
$g_{\alpha\beta}$ \cite{3} \be\label{mm} g_{\alpha\beta}=\Omega^2
\begin{pmatrix}
N_0^2-N_1^2& N_1\\
N_1 & -1
\end{pmatrix},~~~ g^{\alpha\beta}=\frac1{\Omega^2N_0^2}\begin{pmatrix}
1&N_1\\
N_1& N_1^2-N_0^2
\end{pmatrix},~~~ \sqrt{-g}=\Omega^2N_0
\ee with the conformal invariant interval \be\label{me}
ds^2=g_{\alpha\beta}du^{\alpha}du^{\beta}=\Omega^2[N_0^2d\tau^2-
(d\sigma+N_1d\tau)^2], \ee where $N_0$ and $N_1(\tau,\sigma)$ are
known in GR as the lapse function and ``shift vector'',
respectively.

The action (\ref{ss}) after the substitution (\ref{mm}) does not
depend on the conformal factor $\Omega$ and takes the form
\be\label{a4} S=-\frac{\gamma}2
\int_{\tau_1}^{\tau_2}d\tau\int^l_0d\sigma
\left[\frac{(\dot{x}_{\mu}-N_1x'_{\mu})^2}{N_0}-N_0x'{}^2\right],
\ee where $\dot{x}_{\mu}=\partial_{\tau}x_{\mu}$,
$x'_{\mu}=\partial_{\sigma}x_{\mu}$ and
$\dot{x}_{\mu}-N_1x'_{\mu}={D}{}x_{\mu}$ is the covariant
derivative with respect to the metric (\ref{me}). The action
(\ref{a4}), the metric (\ref{me}) and the covariant derivative
$Dx_{\mu}$ are invariant under the ``kinemetric'' transformation
$\tau\rightarrow\tilde{\tau}=f_1(\tau),~
\sigma\rightarrow\tilde{\sigma}=f_2(\tau,\sigma)$. These
 transformations of the differentials
$d\tilde{\tau}=\dot{f}_1(\tau)d\tau$,
$d\tilde{\sigma}=\dot{f}_2(\tau,\sigma)d\tau+{f}_2{}'(\tau,\sigma)d\sigma$
correspond to transformations of the string coordinates
\begin{align}
x_{\mu}{}'(\tau,\sigma)=&~\tilde{x}_{\mu}{}'(\tilde{\tau},\tilde{\sigma})f_2{}
'(\tau,\sigma),\notag\\\label{kintr}
\dot{x}_{\mu}(\tau,\sigma)=&~\dot{\tilde{x}}_{\mu}(\tilde{\tau},\tilde{\sigma})
\dot{f}_1(\tau)+
\tilde{x}_{\mu}{}'(\tilde{\tau},\tilde{\sigma})\dot{f}_2(\tau,\sigma),\\
N_0(\tau,\sigma)=&~\tilde{N}_0(\tilde{\tau},\tilde{\sigma})
\frac{\dot{f}_1(\tau)}{f_2{}'(\tau,\sigma)},~~~
N_1(\tau,\sigma)=\tilde{N}_1(\tilde{\tau},\tilde{\sigma})
\frac{\dot{f}_1}{\tilde{f}_2{}'}+\frac{\dot{f}_2}{\tilde{f}_2{}'},\notag
\end{align}
\be D_{\tau}x_{\mu}(\tau,\sigma)=\dot{f}_1(\tau)D_{\tilde{\tau}}
\tilde{x}_{\mu}(\tilde{\tau},\tilde{\sigma}) \ee

The variation $S$ with respect to $N_0$ and $N_1$ leads to
equation for determination $N_0, N_1$
\begin{align}
\frac{\delta S}{\delta
N_0}=\frac{(Dx_{\mu})^2}{N_0^2}+x'^2=0~\Rightarrow &~
N_0^2=\frac{(\dot{x}x')^2-\dot{x}^2x'^2}{(x'^2)^2},\\
\frac{\delta S}{\delta
N_1}=\frac{(x'{}_{\mu}Dx^{\mu})}{N_0}~\Rightarrow &~
N_1=\frac{(\dot{x}x')}{x'^2},
\end{align}
The substitution of these equation in action (\ref{a4}) converts
it into the standard Nambu--Goto action of the relativistic string
\cite{1} \be S=-\gamma\int_{\tau_1}^{\tau_2}d\tau\int^l_0
d\sigma\sqrt{(\dot{x}x')^2-\dot{x}^2x'^2}.\ee

One can construct the Hamiltonian form of $S$. The conjugate
momenta are determined by

\be p_{\mu}=\frac{\delta S}{\delta
\dot{x}^{\mu}}=\gamma\frac{\dot{x}_{\mu}-N_1x_{\mu}{}'}{N_0}\Rightarrow
\dot{x}_{\mu}=N_0\frac{p_{\mu}}{\gamma}+N_1x_{\mu}{}' \ee and
density of Hamiltonian is obtained by the Legendre transformation
\be\label{ham}
H=p_{\mu}\dot{x}^{\mu}-L=N_0\frac{p^2+\gamma^2x'^2}{2\gamma}+
N_1(px'), \ee then \be S=-\int_{\tau_1}^{\tau_2} d\tau\int_0^l
d\sigma\left[p_{\mu}\dot{x}^{\mu}-H\right]. \ee The secondary
class constraints arises by varying $S$ with respect to $N_0, N_1$
\be\label{con} \frac{\delta S}{\delta
N_0}=\frac{p^2+\gamma^2x'^2}{2\gamma}=0,~~ \frac{\delta S}{\delta
N_1}=(px')=0. \ee The equations of motion take the form
\begin{align}
\frac{\delta S}{\delta
x^{\mu}}=&~\dot{p}_{\mu}-\frac{\partial}{\partial\sigma}(\gamma
N_0x_{\mu}'+N_1p_{\mu})=0\notag,\\
&\hphantom{\frac{\delta S}{\delta
x^{\mu}}=~\dot{p}_{\mu}-\frac{\partial}{\partial\sigma}(\gamma
N_0x_{\mu}'+N_1p_{\mu})=0}\label{ph}\\
 \frac{\delta S}{\delta
p^{\mu}}=&~\dot{x}_{\mu}-N_0\frac{p_{\mu}}{\gamma}+N_1x_{\mu}'=0.\notag
\end{align}

The standard gauge-fixing method is to fix the second class
constraints (orthonormal gauge) $N_0=1, N_1=0.$ In this case the
equations of motion (\ref{ph}) reduce to d'Alambert  ones for
$x_{\mu}$ \be \dot{p}_{\mu}=\gamma x_{\mu}{}'',~~
\dot{x}_{\mu}=\frac{p_{\mu}}{\gamma}\Rightarrow
\ddot{x}_{\mu}-x_{\mu}{}''=0, \ee the conformal interval
(\ref{me}) takes the usual form
$$ ds^2=\Omega^2[d\tau^2-d\sigma^2],$$
but the Hamiltonian (\ref{ham}) in view of the constraint
(\ref{con}) is equal to zero $(H=0)$.

There is another way to introduce evolution parameter as the
object reparameterizations in the theory being adequate to the
initial ``kinemetric'' invariant system and to construct the
non-zero Hamiltonian. We identify this evolution parameter with
the time-like variable of the ``center of mass'' (CM) of a string
defined as the total coordinate \be
X_{\mu}(\tau)=\frac1{l}\int^l_0
x_{\mu}(\tau,\sigma)d\sigma=\langle x_{\mu}\rangle. \ee The
reduction requires to separate the ``center of mass'' variables
before variation of the action (\ref{a4}) which after substitution
\be\label{ix} x_{\mu}=X_{\mu}(\tau)+\xi_{\mu}(\tau,\sigma), ~
x_{\mu}'(\tau,\sigma)=\xi_{\mu}'(\tau,\sigma), \ee \be\label{18}
\int^l_0\xi_{\mu}(\tau,\sigma)=0 \ee takes the form
\begin{multline}
S=\frac{\gamma}{2}\int^{\tau_2}_{\tau_1}d\tau\left\{\dot{X}^2(\tau)\int^l_0
\frac{d\sigma}{N_0(\tau,\sigma)}+2\dot{X}^{\mu}(\tau)\int^l_0
d\sigma \left(\frac{\dot{\xi}_\mu-N_1\xi_{\mu}'}{N_0}\right)+ \right. \\
\left.+\int^l_0
d\sigma\left[\frac{(\dot{\xi}_\mu-N_1\xi_{\mu}')^2}{N_0}-N_0\xi'^2(\tau,\sigma)\right]\right\}.\label{long}
\end{multline}
The usual determination  of the conjugate momenta \be
P_{\mu}(\tau)=\frac{\delta S}{\delta \dot{X}^{\mu}(\tau)}= \gamma
\dot{X}_{\mu}(\tau)\int^l_0\frac{d\sigma}{N_0(\tau,\sigma)}+
\gamma\int^l_0
d\sigma\left(\frac{\dot{\xi}_{\mu}(\tau,\sigma)-N_1\xi_{\mu}'(\tau,\sigma)}{N_0(\tau,\sigma)}\right),
\ee \be \pi_{\mu}(\tau,\sigma)=\frac{\delta
S}{\delta\dot{\xi}^{\mu}(\tau,\sigma)}=\gamma \dot{X}_{\mu}(\tau)
\frac1{N_0(\tau,\sigma)}+\gamma\frac{\dot{\xi}_{\mu}(\tau,\sigma)-N_1\xi_{\mu}'(\tau,\sigma)}
{N_0(\tau,\sigma)}\ee leads to the contradiction because
$P_{\mu}(\tau)$ and $\pi_{\mu}(\tau,\sigma)$ are not independent,
namely $\int^l_0\pi_{\mu}(\tau,\sigma)d\sigma=P_{\mu}(\tau)$.

Thus for the consistent definition  of the momentum of
``center-mass'' $P_{\mu}(\tau)$ and momentum of intrinsic
variables $\pi_{\mu}(\tau,\sigma)$ we have to put strong
constraint  in action (\ref{long}) \be\label{hren} \int
d\sigma\left(\frac{\dot{\xi}_{\mu}(\tau,\sigma)-
N_1\xi_{\mu}'(\tau,\sigma)}{N_0}\right)=0. \ee Then we obtain the
following form of the reduced action: \be S_{\rm
red}=\frac{\gamma}{2}\int d\tau
\left\{\dot{X}^2(\tau)\frac{l}{N(\tau)}+
\int^l_0d\sigma\left[\frac{[\dot{\xi}_{\mu}(\tau,\sigma)-
N_1\xi_{\mu}'(\tau,\sigma)]^2}{N_0(\tau,\sigma)}-
N_0(\tau,\sigma)\xi_{\nu}'{^2}(\tau,\sigma)\right]\right\}, \ee
where $N(\tau)$ is the global lapse function \be\label{glf}
\frac{1}{N(\tau)}=\frac1{l}\int^l_0\frac{d\sigma}{N_0(\tau,\sigma)}=
 \langle N_0^{-1}\rangle.\ee
 For global momenta  we get \be
P_{\mu}(\tau)=\frac{\partial S_{\rm
red}}{\partial\dot{X}^{\mu}(\tau)}=\gamma\dot{X}_{\mu}(\tau)\frac{l}{N(\tau)}
\ee and for the local (intrinsic) momenta  \be
\pi_{\mu}(\tau,\sigma)=\frac{\partial S_{\rm
red}}{\partial\dot{\xi}^{\mu}(\tau,\sigma)}=
\gamma\frac{\dot{\xi}_{\mu}(\tau,\sigma)-
N_1(\tau,\sigma)\xi_{\mu}'(\tau,\sigma)}{N_0(\tau,\sigma)} \ee
with two strong constraints (\ref{18}) and (\ref{hren})
\be\label{a18} \int^l_0\xi_{\mu}(\tau,\sigma)d\sigma=0,~~~~
\int^l_0\pi_{\mu}(\tau,\sigma)d\sigma=0. \ee This separation
conserves the group of the ``kinemetric'' transformation
(\ref{kintr}) and leads to Hamiltonian form of reduced action, in
view of \be
\dot{X}_{\mu}(\tau)=\frac{N(\tau)}{l}\frac{P_{\mu}(\tau)}{\gamma};~~~~~~~
\dot{\xi}_{\mu}(\tau,\sigma)=
\frac{N_0(\tau,\sigma)}{\gamma}\pi_{\mu}(\tau,\sigma)+
N_1(\tau,\sigma)\xi_{\mu}'(\tau,\sigma), \ee  we get \be
H=P_{\mu}\dot{X}^{\mu}(\tau)+\pi_{\mu}\dot{\xi}^{\mu}-L=\frac1{2\gamma}
\left\{\frac{N_0(\tau,\sigma)}{l}P^2(\tau)+
N_0(\tau,\sigma)[\pi^2+\gamma^2\xi'{}^2]+2\gamma
N_1(\pi\xi')\right\}, \ee

\be\label{a}
S=\int^{\tau_2}_{\tau_1}d\tau\left\{P_{\mu}(\tau)\dot{X}^{\mu}(\tau)-
N(\tau)\frac{P^2(\tau)}{2\gamma l}+ \int^l_0 d\sigma
\left[(\pi\dot{\xi})-N_0\frac{\pi^2+\gamma^2\xi'{}^2}{2\gamma}-
N_1(\pi\xi')\right]\right\}. \ee The equation of motion is split
into global one \be \frac{\delta
S}{\delta\dot{X}^{\mu}(\tau)}=\dot{P}_{\mu}(\tau)=0,~~~~~
\frac{\delta S}{\delta
P^{\mu}(\tau)}=\dot{X}_{\mu}(\tau)-{N(\tau)}\frac{P_{\mu}}{\gamma
l}=0 \nonumber \ee and local one \be \frac{\delta
S}{\delta\pi^{\mu}(\tau,\sigma)}=\dot{\xi}_{\mu}
(\tau,\sigma)-\frac{N_0}{\gamma}\pi_{\mu}-N_1\xi_{\mu}'=0,~~~
-\frac{\delta S}{\delta\xi^{\mu}(\tau,\sigma)}=\dot{\pi}_{\mu}-
\frac{\partial}{\partial\sigma}(N_1\pi_{\mu}+\gamma
N_0\xi_{\mu}')=0. \ee The variation of the action (\ref{a}) with
respect to $N_0(\tau)$ results in the constraint \be\label{20}
\frac{\delta S}{\delta
N_0(\tau,\sigma)}=\frac{N^2(\tau)}{N^2_0(\tau,\sigma)}\frac{P^2(\tau)}{2\gamma
l^2}+
\frac{\pi^2(\tau,\sigma)+\gamma^2\xi'{}^2(\tau,\sigma)}{2\gamma}=0,
\ee here it is necessary to take into account that the variation
over the global lapse function (\ref{glf}) leads to \be
\frac{\delta N(\tau)}{\delta
N_0(\tau,\sigma)}=\frac{N^2(\tau)}{N_0^2(\tau,\sigma)}. \ee The
variation of the (\ref{a}) with respect to $N_1(\tau,\sigma)$
results in the constraint for local variables \be
\label{aa20}\frac{\delta S}{\delta
N_1(\tau,\sigma)}=(\pi\xi'(\tau,\sigma))=0. \ee Now we introduce
the Hamiltonian density for local excitations \be\label{a22}
{\cal{H}}(\tau,\sigma)=-\frac{\pi^2(\tau,\sigma)+\gamma^2\xi'^2(\tau,\sigma)}{2\gamma}
\ee and rewrite (\ref{20}) in the form \be\label{a23}
\frac{N(\tau)}{N_0(\tau,\sigma)l}\sqrt{P^2_{\mu}(\tau)}=
\sqrt{\vphantom{P^2_2}2\gamma{\cal H}(\tau,\sigma)}.\ee One
integrates (\ref{a23}) over $\sigma$ and taking into account the
normalization equality (\ref{glf})  \be\label{e}
\frac1l\int^l_0\frac{N(\tau)}{N(\tau,\sigma)}d\sigma=1, \ee it
 leads to global constraint
 \be\label{mst}
M_{\rm st}=\sqrt{P^2_{\mu}(\tau)}=\int^l_0\sqrt{2\gamma{\cal
H}(\tau,\sigma)}d\sigma=l\langle\sqrt{2\gamma{\cal H}}\rangle, \ee
where $M_{\rm st}$ is mass of the string. The local part of the
constraint (\ref{a23}) can be obtained by substitution (\ref{mst})
into (\ref{a23}) \be\label{a26}
\frac{N(\tau)}{N_0(\tau,\sigma)}=\frac{\sqrt{{\cal
H}(\tau,\sigma)}}{\langle\sqrt{{\cal H}}\rangle}. \ee  Finally,
after substitution (\ref{mst}), (\ref{a26}) into action (\ref{a})
we can derive the constraint-shell action, whereas the equation
$N(\tau)P^2(\tau)/2\gamma l=\int N_0(\tau,\sigma){\cal
H}(\tau,\sigma)d\sigma$ \be\label{s} S_{\rm
const-shell}=\int^{\tau_2}_{\tau_1}d\tau\left\{P_{\mu}\dot{X}^{\mu}(\tau)+
\int^l_0d\sigma\left[\pi_{\mu}(\tau,\sigma)\dot{\xi}^{\mu}(\tau,\sigma)-
N_1(\tau,\sigma)\pi_{\mu}(\tau,\sigma)\xi'_{\mu}(\tau,\sigma)\right]\right\}.
\ee Again the variation with respect to $N_1(\tau,\sigma)$ results
in subsidiary condition
\mbox{$(\pi_{\mu}(\tau,\sigma)\xi'{}^{\mu}(\tau,\sigma))=0.$} Now
in the ``center-mass'' coordinate system $P_i(\tau)=0,~~
P_0(\tau)=\int^l_0d\sigma\sqrt{2\gamma{\cal H}(\tau,\sigma)}$ the
action (\ref{s}) takes the form \be\label{a28} S_{\rm
const-shell}= \int^{\tau_2}_{\tau_1}d\tau\int^l_0d\sigma\left\{
\sqrt{2\gamma{\cal
H}(\tau,\sigma)}\dot{X}_0(\tau)+\pi_{\mu}(\tau,\sigma)
\dot{\xi}^{\mu}(\tau,\sigma)\right\}, \ee which describes the
dynamics of a local (intrinsic) canonical variables of a string
with non-zero Hamiltonian because (\ref{a28}) can be rewritten in
the form \be\label{a29} S_{\rm
const-shell}=\int^{x_2}_{x_1}dX_0\int^l_0d\sigma\left\{\pi_{\mu}(X_0,\sigma)
\frac{\partial\xi^{\mu}(X_0,\sigma)}{\partial X_0}+
\sqrt{2\gamma{\cal H}(X_0,\sigma)}\right\}, \ee where
$$2\gamma{\cal H}(X_0,\sigma)=-[\pi^2(X_0,\sigma)+
\gamma^2\xi'{}^2(X_0,\sigma)]$$ and time
$dX_0=\dot{X}_0(\tau)d\tau$ is invariant with respect to
$d\tilde{\tau}=f_1d\tau.$

In the gauge-fixing method, by using the kinemetric
transformation, we have to put \mbox{$N_0(\tau,\sigma)=1,$}
\mbox{$N_1(\tau,\sigma)=0$} (this requirement does not contradict
to Eq. (\ref{e}) in view of Eqs. (\ref{mst}), (\ref{a26})). Then
according to \cite{fr}\be\label{a30} \sqrt{{\cal
H}(\tau,\sigma)}=\frac1l\int^l_0d\sigma\sqrt{{\cal
H}(\tau,\sigma)}=\frac{M_{\rm st}}{\sqrt{2\gamma}~l} \ee  means
that the Hamiltonian $\sqrt{2\gamma{\cal H}(\tau,\sigma)}$ is
constant. In this case the equations for the local variables
obtained by varying the action (\ref{a29}) take the form \be
\frac{\delta S}{\delta \pi^{\mu}(X_0,\sigma)}=\frac{\partial
\xi_{\mu}(X_0,\sigma)}{\partial
X_0}-\frac{\pi_{\mu}(X_0,\sigma)}{\sqrt{-(\pi^2+\gamma^2\xi'{}^2)}}=0,
\ee \be \frac{\delta S}{\delta \xi^{\mu}(X_0,\sigma)}=
-\frac{\partial \pi_{\mu}(X_0,\sigma)}{\partial X_0}+\gamma^2
\frac{\partial}{\partial\sigma}\left[\frac{
\xi'_{\mu}(X_0,\sigma)}{\sqrt{-(\pi^2+\gamma^2\xi'{}^2)}}\right]=0.
\ee If we put according to Eq. (\ref{a30})
$\sqrt{-(\pi^2+\gamma^2\xi'{}^2)}=\gamma,~~ (M_{\rm st}=\gamma
l)$, then it  leads again to d'Alambert equation for
$\xi_{\mu}(X_0,\sigma)$ \be \frac{\partial^2
\xi_\mu{(X_0,\sigma)}}{\partial X^2_0}=\frac{\partial^2
\xi_\mu{(X_0,\sigma)}}{\partial \sigma^2}. \ee The general
solution of these equations in class of functions (\ref{a18}) with
boundary conditions for the open string
$\xi'{}_{\mu}(X_0,0)=\xi'{}_{\mu}(X_0,l)=0$ is given as usually by
the Fourier series
\begin{align}
\xi_{\mu}(X_0,\sigma)=&\frac1{2\sqrt{\gamma}}[\Psi_{\mu}(X_0+\sigma)+
\Psi_{\mu}(X_0-\sigma)],\notag\\
\xi'{}_{\mu}(X_0,\sigma)=&\frac1{2\sqrt{\gamma}}[\Psi'{}_{\mu}(X_0+\sigma)-
\Psi'{}_{\mu}(X_0-\sigma)],\label{a47}\\
\pi_{\mu}(X_0,\sigma)=&\gamma\frac{\partial\xi_{\mu}(X_0,\sigma)}{\partial
X_0}=\frac{\sqrt{\gamma}}{2}[\Psi'{}_{\mu}(X_0+\sigma)+
\Psi'{}_{\mu}(X_0-\sigma)],\notag
\end{align}
where \be\label{a32} \Psi_{\mu}(z)=i\sum_{n\neq
0}\frac{\alpha_{n\mu}}{n}e^{-\frac{i\pi n}{l}z},~~
\Psi'{}_{\mu}(z)=\frac{\pi}l\sum_{n\neq
0}\alpha_{n\mu}e^{-\frac{i\pi n}{l}z} \ee it is very important to
stress that (\ref{a32}) does not contain zero harmonics $(n\neq
0)$). The substitution of $\xi_{\mu}$ and $\pi_{\mu}$ in this form
into (\ref{a22}) and taken into consideration (\ref{a30}) leads to
density of Hamiltonian \be\label{a33} {\cal
H}=-\frac14\left[\Psi'{}^2_{\mu}(X_0+\sigma)+
\Psi'{}^2_{\mu}(X_0-\sigma)\right]=\frac{M^2_{\rm st}}{2\gamma
l^2}. \ee From the constraint (\ref{aa20}) $(\pi\xi')=0$ in terms
of the vector $\Psi'{}_{\mu}$ (\ref{a47}) we obtain \be\label{a34}
\left(\pi_{\mu}(X_0,\sigma)\xi'{}^{\mu}(X_0,\sigma)\right)=
\frac14\left[\Psi'{}_{\mu}^2(X_0+\sigma)-\Psi'{}_{\mu}^2(X_0-\sigma)\right]=0,
\ee then from (\ref{a33}) and (\ref{a34}) finally we get
\be\label{eq}
\Psi_{\mu}'{}^2(X_0+\sigma)=\Psi_{\mu}'{}^2(X_0-\sigma)=-\frac{M_{\rm
st}^2}{l^2\gamma}.\ee It means that $\Psi_{\mu}'(z)$ is the
modulo-constant space-like vector and in terms of its
representation (\ref{a32}) the equalities (\ref{eq}) can be
rewritten \be\label{a36}
-\Psi_{\mu}'{}^2(z)=\frac{\pi^2}{l^2}\sum^{\infty}_{k=-\infty}
L_ke^{-i\frac{\pi k}{l}z}=\frac{M^2_{\rm st}}{l^2 \gamma},\ee
where
$$L_k=-\sum_{n\neq k,0}\alpha_{n\mu}\alpha^{\mu}_{k-n},~~~~
L_k^{*}=L_{-k}~.$$ Now we can see that the zero harmonic of this
constraint determines the mass of a string \be M_{\rm
st}^2=\pi^2\gamma L_0=-\pi\gamma
\sum_{n\neq0}\alpha_{n\mu}\alpha^{\mu}_{-k},~~~~
\alpha^{\mu}_{-k}=\alpha^{*}_k{}^{\mu} \ee and coincides with
standard definition in string theory \cite{1}, however the
non-zero harmonics of constraint (\ref{a36}) \be\label{a37}
L_{k\neq0}=-\sum_{n\neq0,k}\alpha_{nk}\alpha^{\mu}_{k-n}=0 \ee
strongly differ from the standard theory because they do not
depend on the global motion(do not depend on $P_{\mu}$) and do not
contain the interference term $P_{\mu}\Psi^{\mu}{}'$, because of
our constraint (\ref{eq}) we can rewrite \be
l^2\gamma\Psi_{\mu}'{}^2(z)+P^2_{\mu}=0 \ee instead standard one
\cite{1} \be (P_{\mu}+l\sqrt{\gamma}\Psi_{\mu}'(z))^2=0. \ee
Therefore diffeo-invariant approach coincides with the
R{\"o}hrlich one \cite{4}, which is based on the gauge condition
$P_{\mu}\xi^{\mu}=0,~~~ P_{\mu}\pi^{\mu}=0 \Rightarrow
P_{\mu}\alpha_n^{\mu}=0,~~~ n\neq0$. One use of that condition for
eliminating the time components $\xi_0, \pi_0$ being constructed
in the frame of reference $(P_i=0)$ leads to formula
(\ref{a47})--(\ref{a37}), where $\Psi'_{0}$ and $\alpha_{n0}$ are
equal to zero.

\section{Conclusions}

   The standard approach to a relativistic string
   is  not invariant with respect to
   the time coordinate transformations. It is the problem
 as, in diffeo-invariant theories, all  observable quantities
  should be the  diffeo-variant ones.

 We propose here to solve this problem as in \cite{bpp}, where
  the reference frame
   is redefined by pointing out
  diffeo-invariant homogeneous {\it``time-like variable''}
  in accordance with the dimension of the diffeomorphism group.

  Just this  choice
   of an evolution parameter
  simplifies the Hamiltonian
  equations and
   leads to exact resolution of the energy constraint
 with respect to  the canonical momentum of the evolution parameter.
 We have shown that this diffeo-invariant approach to a
 relativistic string corresponds to the R{\"o}hrlich spectrum
 \cite{4}.


\vspace{1cm}

{\bf Acknowledgement}

\medskip

The authors are grateful to   A.V. Efremov,
 E.A. Kuraev,   V.V. Nesterenko,
 for interesting and critical
discussions.


\begin{thebibliography}{}


\bibitem{1}
B.M.~Barbashov, V.V.~Nesterenko, {\it Introduction in the
Relativistic String Theory} (World Scientific, 1990).
\bibitem{H} D. Hilbert, 
 Nachrichten von der K\"on. Ges. der Wissenschaften zu G\"ottingen,
 Math.-Phys. Kl., Heft 3,  395 (1915).
\bibitem{H24} D.~Hilbert, 
 Math. Annalen.  {\bf 92}, 1 (1924).

\bibitem{d1986} V.V.~Nesterenko, {\it``About Interpretation of
the   Noether Identities''} Preprint JINR - P2-86-284, Dubna,
1986.

\bibitem{Noter} E.  Noether, G\"ottinger  Nachrichten,
 Math.-Phys. Kl., {\bf 2}, 235 (1918).

\bibitem{vlad}
A.L.~Zelmanov, Dokl. AN USSR {\bf 107},  315 (1956);
A.L.~Zelmanov, Dokl. AN USSR {\bf 209},  822 (1973);
Yu.S.~Vladimirov, {\it Reference Frames in Theory of Gravitation},
Moscow, Energoizdat, 1982 [in Russian].



\bibitem{bpp}
B.M.~Barbashov, V.N.~Pervushin, and D.V.~Proskurin, {Theoretical
and Mathematical Physics}, {\bf 132}, 1045 (2002).

\bibitem{dir}
P.A.M.~Dirac, Proc.Roy.Soc. {\bf A 246},  333 (1958);
P.A.M.~Dirac, Phys. Rev. {\bf 114}, 924 (1959);  R.~Arnowitt,
S.~Deser and C.W.~Misner, Phys. Rev. {\bf 117},   1595 (1960).

\bibitem{WDW}
  J.A.~Wheeler, in \newblock{\em Batelle Recontres: 1967,  Lectures in Mathematics
and Physics,} edited by C. DeWitt and J.A. Wheeler , New York, 1968,
p.~242;
 B.C.~DeWitt,  Phys. Rev. {\bf 160}, 1113 (1967).

\bibitem{fr}
   B.M.~Barbashov and N.A.~Chernikov,
{\it Classical Dynamics of Relativistic String} [in Russian],
Preprint JINR P2-7852,
 Dubna, 1974;
 A.G.~Reiman and L.D.~Faddeev,
Vestn. Leningr. Gos. Univ., No.1,  138 (1975) [in Russian].


\bibitem{2}
L. Brink, P. Di Vecchia, P. Howe, Phys. Lett {\bf B 65}, 47
(1976).

\bibitem{3}
 B.M.~Barbashov and V.N.~Pervushin,
 {Theor. Math. Phys.}
 {\bf 127} 483 (2001); [hep-th/0005140].

\bibitem{4}
F.~R\"ohrlich, Phys. Rev. Lett. {\bf 34}, 842 (1975); Nucl. Phys.
{\bf B 112}, 177 (1976); Ann. Phys. {\bf 117}, 292 (1979).



\bibitem{fock29} V.A.~Fock, Zs. f. Phys. {\bf 57},
261 (1929).



%
%




%
%



\end{thebibliography}
\end{document}